\documentstyle[epsfig,longtable]{aipproc}
\begin{document}
\title{A New Measurement of the Energy Dependence of Nuclear Transparency 
            for Large Momentum Transfer $^{12}C(p,2p)$ Scattering}
 
\author{
\centering
A. Leksanov$^a$, J. Alster$^b$, G. Asryan$^c$, Y. Averichev$^{h}$,
\newline
  D. Barton$^{d}$,
 V. Baturin$^{a,e}$, N. Bukhtojarova$^{d,e}$, 
  A. Carroll$^c$,
\newline
 A. Schetkovsky$^{a,e}$,
  S. Heppelmann$^a$, T. Kawabata$^f$, A. Malki$^b$,
\newline
  Y. Makdisi$^c$, E. Minina$^a$,
I. Navon$^b$, H. Nicholson$^g$,
\newline
 A. Ogawa$^a$, Y. Panebratsev$^{h}$,
  E. Piasetzky$^b$,
 S. Shimanskiy$^{h}$,
\newline
  A. Tang$^i$, J.W. Watson$^i$, H. Yoshida$^f$, D. Zhalov$^a$
}
\address{$^a$Physics Department, Pennsylvania State University, University Park, PA 16801, USA \\
$^b$School of Physics and Astronomy, Sackler Faculty of Exact  Sciences, Tel Aviv University, Ramat Aviv 69978, Israel \\
$^c$Yerevan Physics Institute, Yerevan 375036, Armenia \\
$^d$Collider-Accelerator Department, Brookhaven National Laboratory, Upton, NY 11973, USA \\
$^e$Petersburg Nuclear Physics Institute, Gatchina, St. Petersburg 188350, Russia \\
$^f$Department of Physics, Kyoto University, Sakyoku, Kyoto, 606-8502, Japan \\
$^g$Department of Physics, Mount Holyoke College, South Hadley, MA 01075, USA \\
$^h$J.I.N.R., Dubna, 141980, Russia\\
$^i$Department of Physics, Kent State University, Kent, OH 44242, USA\\
}

\maketitle

\begin{abstract}
    We present a new measurement of the energy dependence of nuclear 
    transparency from AGS experiment E850, performed using the
    EVA solenoidal spectrometer, upgraded since 1995. Using a secondary 
    beam from the AGS accelerator, we simultaneously measured $pp$ 
    elastic scattering from hydrogen and $(p,2p)$
    quasi-elastic scattering in carbon at incoming momenta of
    5.9, 8.0, 9.0, 11.7 and 14.4 GeV/c.  This incident momentum range 
    corresponds to a $Q^{2}$ region between 4.8 and 12.7 (GeV/c)$^{2}$.
    The detector allowed us to do a complete kinematic analysis 
    for the center-of-mass polar angles in the range $85^{\circ}-90^{\circ}$.
    We report on the measured variation of the nuclear transparency 
    with energy and compare the new results with previous measurements. 

\end{abstract}

Color Transparency(CT) is the predicted reduction
in the initial and final $\,$state interactions, which may 
take place when a large transverse momentum($p_{t}$),
quasi-exclusive scattering, involving hadrons,
occurs in nuclear matter. We study nuclear
transparency for $(p,2p)$ quasi-elastic scattering. It can be 
informally defined as 
\begin{equation}
	T_{r}=\frac{\frac{d\sigma}{dt}(\mbox{$pp$ quasi-elastic in nucleus})}{Z\frac{d\sigma}{dt}(\mbox{$pp$ elastic in hydrogen})}\,\,\,\,.
\label{eq:trdef}
\end{equation}
Theoretical predictions for the variations of this quantity with energy are
very model dependent. For example, Glauber-based calculations show no
energy dependence above 5 GeV with a value around thirty percent, 
whereas QCD based models predict an increase of $T_{r}$ with energy, reaching 
eventually an asymptotic value of unity \cite{farrar-1}. 
Interpretations of fluctuating energy dependence of the transparency 
\cite{carrol-1} at the values of $Q^{2}$ above 10 (Gev/c)$^{2}$ have been
presented in the models described in Refs. \cite{rals-1} and \cite{brod-1}.

Many widely accepted pictures of CT dynamics
are based on the assumption that selection of small-size 
Fock state configurations occurs in certain hard scattering processes
and that these small configurations have a reduced cross-section for
reinteraction with the surrounding nucleus.
The protons remain in these configurations during a time period which 
is different for different energies and large on the scale of nuclear
processes \cite{strik-1}.


We performed our measurement using the EVA apparatus located
in the C1 line of the AGS at BNL. 
Two differential Cerenkov counters were used for the
incident particle identification. The beam flux measurement was
performed using a scintillating-fiber beam hodoscope located
in the upstream part of the apparatus.
We used solid carbon and polyethylene targets, whose dimensions were 
5.1~cm~$ \times$~5.1~cm~$\times$~6.6~cm. The AGS provided a secondary beam with  
approximately 1-2$\times10^{7}$ particles per 2 second spill and about
20 spills per minute.
EVA is a magnetic
spectrometer with tracking capabilities.
It was built around a solenoidal superconducting magnet
which produced
nearly uniform field in the beam direction allowing for charged particle
transverse momentum($p_{t}$) measurements. It further allowed us to reject 
low $p_{t}$ tracks at the trigger level.

The tracking with this apparatus used an array of four cylindrical
straw tube drift chambers with charge division, 
providing us with 3D position information.
The chambers were used both for triggering and 
for final track reconstruction.
The signals from two fan shaped hodoscopes of scintillating counters
served as input to the first level trigger, which started the digitization
of the event \cite{wu-1},\cite{mardor-1},
\cite{mardor-3},\cite{durrant-1},\cite{lek-1}.

The variable we present in this paper is the {\it transparency
ratio} $T_{CH}$, which we define to be 
the ratio of $C(p,2p)$ quasi-elastic(QE) to $pp$ elastic 
cross-section divided by the $Z$ of the nucleus(6 in our case) for 
a restricted region of kinematics near that of $pp$ elastic scattering.
Within the
impulse approximation this quantity is related to nuclear transparency via
\begin{equation}
T_{CH}=\frac{C}{ZH}=T_{r}\int^{\alpha_{2}}_{\alpha_{1}}\frac{d\alpha}{\alpha}\int\int d\vec{p}_{Ft}n(\alpha,\vec{p}_{Ft})\frac{\frac{d\sigma}{dt}(s)}{\frac{d\sigma}{dt}(s_{0})}\,\,\,.
\label{eq:tr_rat}
\end{equation}
Here $n(\alpha,\vec{p}_{Ft})$ is the momentum distribution of the 
target proton in the nucleus,
$\alpha = \frac{E_{F}-p_{Fz}}{m} \simeq \frac{s}{s_{0}}$ near $\alpha = 1$,
$s$ and $s_{0}$ are the Mandelstam variables for $C$ and $H$ 
events correspondingly,
$\vec{p}_{Ft} = (p_{Fx},p_{Fy})$ is the transverse component of the target nucleon momentum,
$\frac{d \sigma}{dt}(s)$ is the differential $pp$ cross-section for 
the process inside the nucleus, $\frac{d \sigma}{dt}(s_{0})$ - for the 
hydrogen.

In the presentation of these measurements, we attempt to 
limit the influence of incomplete knowledge of the spectral
function. This was accomplished by applying an unrestrictive cut on
transverse Fermi momentum but a tight cut on longitudinal 
nuclear motion so that  
$\frac{d \sigma}{dt}(s) \simeq \frac{d \sigma}{dt}(s_{0})$,
neglecting small variations in cross-section.

Our apparatus was capable of measuring all three components
of momenta of both outgoing particles. Therefore the kinematic
variables of interest were calculated from the data. The following 
cuts ensured that the conditions were satisfied:
$\vert p_{Fy} \vert < 0.3$ GeV/c, $ \vert p_{Fx} \vert < 0.5$ GeV/c and
$ 0.95 < \alpha_{0} <1.05$. The variable $\alpha_{0}$, defined as
\begin{equation}
\alpha_{0} = 1 - \frac{ 2p \cos{\frac{\theta_{1}-\theta_{2}}{2}} \cos{\frac{\theta_{1}+\theta_{2}}{2}}-p_{inc \, \, z}}{m}
\end{equation}
with $p=\sqrt{(\frac{E_{beam}+m}{2})^{2}-m^{2}}$, is an 
analog of $\alpha$, which was defined to be independent
of momentum variables and varies only with the opening
angle in the Lab frame. The value of 
$\vert \alpha - \alpha_{0} \vert$ in the region of interest is well below 
one percent
\cite{mardor-1}, \cite{mardor-3}. 

To extract the number of events from $C$ and $CH_{2}$ targets,
we used the missing mass variable defined as:
\begin{eqnarray}
\label{mmdef}
M_{miss}^{2}=E_{miss}^{2}-\vec{p}_{F}^{\,\,2}\,\,\,.
\end{eqnarray}
As it follows from equation\ (\ref{mmdef}), $M_{miss}^{2}$ distribution for
the reconstructed QE events should display a peak around zero.
From the events that satisfy the cuts we select
candidates for QE by demanding exactly two charged tracks in the 
final state. We also select events with more then two tracks 
to model the behavior of the background. The signal is then extracted
from these distributions after background subtraction.

The hydrogen signal was measured as the normalized difference between the yields from the 
$CH_{2}$ and $C$ targets. Then the transparency ratio can be defined as
\begin{equation}
T_{CH}=\frac{\beta\gamma C}{3(CH_{2}-\beta\gamma C)}\,\,\,.
\label{eq:tch}
\end{equation}
Here $C$ and $CH_{2}$ are the numbers of extracted events.
$\beta = 0.4535 $ is the ratio of number of $CH_{2}$ molecules to the number of $C$
atoms in the corresponding targets, $\gamma$ is the beam normalization factor $\gamma = \frac{beam_{C}}{beam_{CH_{2}}}$. The beam normalization factor is determined from the 
flux measurements using scaler readings and comparing yields for the 
same target positions and different target configurations. 

\begin{figure}[h!]
\centerline{\epsfig{figure=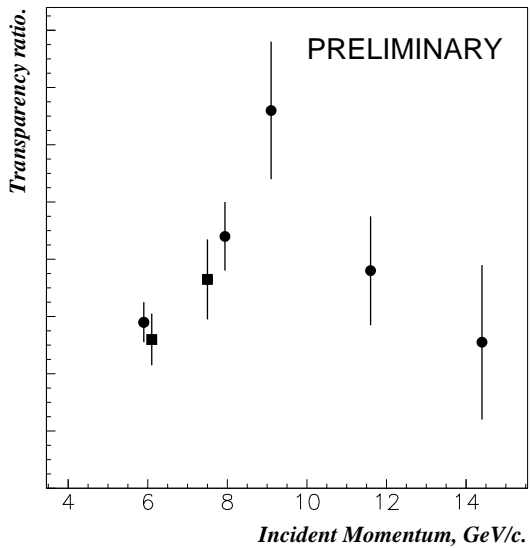, height=8cm,width=8cm}}
\vspace{10pt}
\caption{The measured dependence of the transparency ratio, $T_{CH}$, on the
incident momentum, $P_{inc}$. Boxes represent 1994 results and circles - 1998.
No systematic errors are included.}
\label{transp.fig}
\end{figure}

We performed our measurements  at incoming momenta of
$5.9$ and $7.5$ GeV/c in 1994 \cite{mardor-2} and 
$5.9$, $8.0$, $9.0$, $11.7$ and $14.4$ GeV/c in 1998.
These corresponded to the following values of momentum transfer $Q^{2}$:
$4.8$ and $6.2$ (GeV/c)$^2$ in 1994 and 
$4.8$, $6.6$, $7.6$, $10.1$ and $12.7$  (GeV/c)$^2$ in 1998.
The preliminary result of our experiment, 
the energy dependence of the transparency
ratio, integrated over the center-of-mass scattering angle in the region 
$85^{\circ}-90^{\circ}$, is presented in the figure ~\ref{transp.fig}.  
We can draw the following conclusions from it. First, the ratio definitely 
varies with energy with strong deviations from Glauber calculations.  Second, we clearly see
the rise of the ratio for the incident momenta below 10 GeV/c.
Third, we also observe the fall of the transparency above 10 GeV/c.
Finally, our results seem to be in a good agreement with 
the energy dependence reported in 1988 by
E834 ~\cite{carrol-1}

\end{document}